\documentclass{article}
\usepackage{times}
\usepackage{amsfonts}
\begin{document}
\title{Gerbes, Quantum Mechanics and Gravity}
\author{Jos\'e M. Isidro\\
Instituto de F\'{\i}sica Corpuscular (CSIC--UVEG)\\
Apartado de Correos 22085, Valencia 46071, Spain\\
{\tt jmisidro@ific.uv.es}\\
and\\
Max--Planck--Institut f\"ur Gravitationsphysik, 
\\Albert--Einstein--Institut, \\
D--14476 Golm, Germany\\
{\tt isidro@aei.mpg.de}}

\maketitle

\begin{abstract}

\noindent

\end{abstract}
\noindent
We prove that invariance of a quantum theory under the semiclassical {\it vs.}\/ strong--quantum duality $S/\hbar\longleftrightarrow\hbar/S$, where $S$ is the classical action, is equivalent to noncommutativity (of the Heisenberg--algebra type) of the coordinates of the space on which $S$ is defined. We place these facts in correspondence with gerbes and Neveu--Schwarz $B$--fields and discuss their implications for a quantum theory of gravity. Feynman's propagator turns out to be closely related to the trivialisation of a gerbe on configuration space.


\section{Introduction}\label{intt}

Let $\mathbb{M}$ be an $n$--dimensional spacetime manifold endowed with the the metric tensor $g_{\mu\nu}$. Let $x^{\mu}$, $\mu=1,\ldots, n$, be local coordinates on $\mathbb{M}$. The possibility of measuring the infinitesimal distance 
\begin{equation}
{\rm d}s^2=g_{\mu\nu}{\rm d}x^{\mu}{\rm d}x^{\nu}
\label{kkaa}
\end{equation}
between two points on $\mathbb{M}$ rests on the assumption that the corresponding coordinates can be simultaneously measured with infinite accuracy, so one can have
\begin{equation}
\Delta x^{\mu}=0, 
\label{rda}
\end{equation}
simultaneously for all $\mu=1,\ldots, n$. In quantum--mechanical language one would recast this assumption as 
\begin{equation}
[x^{\mu},x^{\nu}]=x^{\mu}x^{\nu}-x^{\nu}x^{\mu}=0.
\label{ccss}
\end{equation}
The vanishing of the above commutator expresses two alternative, though essentially equivalent, statements, one of physical content, the other geometrical. Physically it expresses the absence of magnetic $B$--fields across the $\mu,\nu$ directions \cite{LANDAU}. Geometrically it expresses the fact that the multiplication law on the algebra of functions on spacetime $\mathbb{M}$ is commutative.

Modern theories of quantum gravity, as well as string theory, all share the common feature that a minimal length scale, the {\it Planck length}\/ $L_P$, exists on spacetime, 
\begin{equation}
\Delta x^{\mu}\geq L_P,
\label{bbcc}
\end{equation}
so $L_P$ effectively becomes the shortest possible distance,
\begin{equation}
{\rm d}s^2\geq L_P^2. 
\label{rrmm}
\end{equation}
This coarse graining of a spacetime continuum $\mathbb{M}$ can be mimicked, in noncommutative geometry \cite{NCG}, by noncommuting {\it operator coordinates}\/ $\hat x^{\mu}$ acting as Hermitean operators on Hilbert space $\mathbb{H}$. The $\hat x^{\mu}$ satisfy
\begin{equation}
[\hat x^{\mu},\hat x^{\nu}]={\rm i}a\theta^{\mu\nu},
\label{cghp}
\end{equation}
with $\theta^{\mu\nu}$ a constant, real, dimensionless antisymmetric tensor. Here $a>0$ is a fundamental area scale, such that
\begin{equation}
\lim_{a\to 0}[\hat x^{\mu},\hat x^{\nu}]=0.
\label{rktdpk}
\end{equation}
Moreover, in the limit $a\to 0$, one can identify (possibly up to some singular renormalisation factor $Z$) the operator $\hat x^{\mu}$ on $\mathbb{H}$ with the function $x^{\mu}$ on $\mathbb{M}$. Since the Heisenberg uncertainty relations corresponding to (\ref{cghp}) imply
\begin{equation}
\Delta\hat x^{\mu}\Delta\hat x^{\nu}\geq \frac{a}{2}\vert\theta^{\mu\nu}\vert, 
\label{rkbr}
\end{equation}
the above statement concerning the coarse graining of $\mathbb{M}$ follows. Up to possible numerical factors one can therefore set
\begin{equation}
a=L_P^2.
\label{rzf}
\end{equation}

It has been argued \cite{PADUNO} that the existence of a fundamental length scale $L_P$ on $\mathbb{M}$ implies modifying the spacetime metric according to the rule
\begin{equation}
{\rm d}s^2\longrightarrow{\rm d}s^2+L_P^2,
\label{rkcak}
\end{equation}
so $L_P$ effectively becomes the shortest possible distance. One can also prove \cite{PADUNO} that modifying the spacetime interval according to (\ref{rkcak}) is equivalent to requiring invariance of a field theory under the following exchange of short and long distances:
\begin{equation}
{\rm d}s\longleftrightarrow\frac{L_P^2}{{\rm d}s}.
\label{paddy}
\end{equation}

On the other hand, we have in ref. \cite{ME} shown that the existence of a minimal length scale $L_P$ is equivalent to the exchange
\begin{equation}
\frac{S}{\hbar}\longleftrightarrow\frac{\hbar}{S}
\label{rchp}
\end{equation}
in Feynman's exponential of the action integral $S$
\begin{equation}
\exp\left({\rm i}\,\frac{S}{\hbar}\right),
\label{bbmk}
\end{equation}
so the latter becomes
\begin{equation}
\exp\left({\rm i}\,\frac{\hbar}{S}\right).
\label{bbmc}
\end{equation}
Since the equations of motion that follow from the variation of $S/\hbar$ are the same as those derived from the variation of $\hbar/S$, classically there is no difference between $S/\hbar$ and $\hbar/S$.  We will refer to the exchange (\ref{rchp}) as {\it semiclassical vs. strong--quantum duality}. This simple $\mathbb{Z}_2$--transformation has been extended \cite{ME}, in the sense of group--theoretic extensions \cite{AZCA}, to larger duality groups $G$ such as ${\rm SL}\,(2,\mathbb{Z})$, ${\rm SL}\,(2,\mathbb{R})$  and ${\rm SL}\,(2,\mathbb{C})$.

It is the purpose of this letter to examine the relation between the noncommutativity (\ref{cghp}) and the semiclassical {\it vs.}\/ strong--quantum duality (\ref{rchp}). We will prove that eqns. (\ref{cghp}) and (\ref{rchp}) are equivalent: whenever the one holds, so does the other, and viceversa. For the proof we will use the geometric language of gerbes \cite{GERBES}.

That eqns. (\ref{cghp}) and (\ref{rchp}) should somehow be equivalent can be arrived at reasoning as follows. Eqn. (\ref{rzf}) implies that if we have a length scale $L_P$ then we have an area scale $a=L_P^2$, and conversely. The latter appears on the right--hand side of (\ref{cghp}). Building on previous work \cite{PADUNO}, the equivalence between the existence of a length scale $L_P$ and the duality (\ref{rchp}) has been proved in ref. \cite{ME}. Hence one can expect the commutation relations (\ref{cghp}) to follow from the duality (\ref{rchp}), and viceversa.

An instance of the semiclassical {\it vs.}\/ strong--quantum duality (\ref{rchp}) has appeared under a different, though essentially equivalent, guise, in ref. \cite{PINZULSTERN}. In this latter paper the reparametrisation symmetry of a relativistic free particle has been exploited to impose a gauge condition which, upon quantisation, implies spacetime noncommutativity. Then an algebraic map from this gauge back to the standard commuting gauge has been shown to exist such that the classical Poisson algebra, and the resulting quantum theory, are identical in the two gauges. Other recent works on gerbes, duality, noncommutative quantum mechanics and related topics are refs. \cite{RECENT, VARIOUS}. Looking beyond, eqn. (\ref{rchp}) presumably has close cousins in other known duality transformations in fields, strings and branes \cite{VAFA}, T--duality \cite{PORRATI} being one example.

\section{Basics in gerbes}\label{rman}

A comprehensive treatment of gerbes can be found in ref. \cite{GERBES};  nice reviews are refs. \cite{ORLANDO} and \cite{PICKEN}. 

It is well known that a unitary line bundle on a base manifold $\mathbb{B}$ is a 1--cocycle $\lambda\in H^1\left(\mathbb{B}, C^{\infty}({\rm U}(1))\right)$. The latter is the first \v Cech cohomology group of $\mathbb{B}$ with coefficients in the sheaf of germs of smooth, U(1)--valued functions. Let $\left\{U_{\alpha}\right\}$ be a good cover of $\mathbb{B}$ by open sets $U_{\alpha}$. Then the bundle is determined by a collection of U(1)--valued transition functions defined on each 2--fold overlap
\begin{equation}
\lambda_{\alpha_1\alpha_2}:U_{\alpha_1}\cap U_{\alpha_2}\longrightarrow {\rm U}(1)
\label{marsalis}
\end{equation}
satisfying 
\begin{equation}
\lambda_{\alpha_2\alpha_1}=\lambda^{-1}_{\alpha_1\alpha_2}, 
\label{winton}
\end{equation}
as well as the 1--cocycle condition
\begin{equation}
\lambda_{\alpha_1\alpha_2}\lambda_{\alpha_2\alpha_3}\lambda_{\alpha_3\alpha_1}=1\quad {\rm on}\quad U_{\alpha_1}\cap U_{\alpha_2}\cap U_{\alpha_3}.
\label{ptbb}
\end{equation}

A {\it gerbe}\/  is defined as a 2--cocycle $g\in H^2\left(\mathbb{B}, C^{\infty}({\rm U}(1))\right)$. This means that we have a collection $\{g_{\alpha_1\alpha_2\alpha_3}\}$ of maps defined on each 3--fold overlap on $\mathbb{B}$
\begin{equation}
g_{\alpha_1\alpha_2\alpha_3}:U_{\alpha_1}\cap U_{\alpha_2}\cap U_{\alpha_3}\longrightarrow {\rm U}(1)
\label{bbmkmd}
\end{equation}
satisfying
\begin{equation}
g_{\alpha_1\alpha_2\alpha_3}=g^{-1}_{\alpha_2\alpha_1\alpha_3}=g^{-1}_{\alpha_1\alpha_3\alpha_2}=g^{-1}_{\alpha_3\alpha_2\alpha_1},
\label{ktdpkbb}
\end{equation}
as well as the 2--cocycle condition
\begin{equation}
g_{\alpha_2\alpha_3\alpha_4}\,g^{-1}_{\alpha_1\alpha_3\alpha_4}\,g_{\alpha_1\alpha_2\alpha_4}\,g^{-1}_{\alpha_1\alpha_2\alpha_3}=1\quad {\rm on}\quad U_{\alpha_1}\cap U_{\alpha_2}\cap U_{\alpha_3}\cap U_{\alpha_4}.
\label{knptbb}
\end{equation}
Now $g$ is a 2--coboundary in \v Cech cohomology whenever it holds that
\begin{equation}
g_{\alpha_1\alpha_2\alpha_3}=\tau_{\alpha_1\alpha_2}\tau_{\alpha_2\alpha_3}\tau_{\alpha_3\alpha_1}
\label{lag}
\end{equation}
for a certain collection $\left\{\tau_{\alpha_1\alpha_2}\right\}$ of U(1)--valued functions $\tau_{\alpha_1\alpha_2}$ on $U_{\alpha_1}\cap U_{\alpha_2}$ such that $\tau_{\alpha_2\alpha_1}=\tau^{-1}_{\alpha_1\alpha_2}$. The collection $\left\{\tau_{\alpha_1\alpha_2}\right\}$ is called a {\it trivialisation}\/ of the gerbe. One can prove that over any given open set $U_{\alpha}$ of the cover $\left\{U_{\alpha}\right\}$ there always exists a trivialisation of the gerbe. Moreover, any two trivialisations $\left\{\tau_{\alpha_1\alpha_2}\right\}$, $\{\tau'_{\alpha_1\alpha_2}\}$ differ by a unitary line bundle. This is so because the quotient $\tau'_{\alpha_1\alpha_2}/\tau_{\alpha_1\alpha_2}$ satisfies the 1--cocycle condition (\ref{ptbb}). A gerbe, however, does not qualify as a manifold, since the difference between any two trivialisations is not a transition function, but a line bundle. To compare with fibre bundles, the total space of a bundle is always a manifold, any two local trivialisations differing by a transition function.

On a gerbe specified by the 2--cocycle $g_{\alpha_1\alpha_2\alpha_3}$, a connection is specified by forms $A, B, H$ satisfying
\begin{eqnarray}
H\vert_{U_{\alpha}}&=&{\rm d}B_{\alpha}\\
B_{\alpha_2}-B_{\alpha_1}&=&{\rm d}A_{\alpha_1\alpha_2}\\
A_{\alpha_1\alpha_2}+A_{\alpha_2\alpha_3}+A_{\alpha_3\alpha_1}&=&g^{-1}_{\alpha_1\alpha_2\alpha_3}{\rm d}g_{\alpha_1\alpha_2\alpha_3}.
\label{ktpyy}
\end{eqnarray}
The 3--form $H$ is the curvature of the gerbe connection. The latter is called {\it flat}\/ if $H=0$.

\section{A gerbe on configuration space}\label{chpbbln}

\subsection{The trivialisation}\label{vtr}

Let an action integral $S$ be given for a point particle on spacetime $\mathbb{M}$. Let us further assume that the latter factorises, at least locally, as the product of the time axis $\mathbb{R}$ and a configuration space $\mathbb{F}$. Coordinates $x_{(\alpha)}^{\mu}$ on the local chart labelled by ${\alpha}$ therefore decompose as $(t_{\alpha}, q_{\alpha}^{j})$, with $j=1$, $\ldots$, $n-1$. This latter index will be suppressed in what follows. Let any two points $q_{\alpha_1}$, $q_{\alpha_2}$ be given on $\mathbb{F}$, with local charts $U_{\alpha_1}$, $U_{\alpha_2}$ centred around them. Moreover, let $\mathbb{L}_{\alpha_1\alpha_2}$ be an oriented path connecting $q_{\alpha_1}$ to $q_{\alpha_2}$ as time runs from $t_{\alpha_1}$ to $t_{\alpha_2}$. We define $\tilde a_{\alpha_1\alpha_2}$ as the following functional integral over all such trajectories $\mathbb{L}_{\alpha_1\alpha_2}$:
\begin{equation}
\tilde a_{\alpha_1\alpha_2}\sim\int {\rm D}\mathbb{L}_{\alpha_1\alpha_2}\exp\left(\frac{{\rm i}}{\hbar}S(\mathbb{L}_{\alpha_1\alpha_2})\right).
\label{bbkkp}
\end{equation}
Throughout this paper, the $\sim$ sign  will stand for {\it proportionality}: path integrals are defined up to some (usually divergent) normalisation. However all such normalisation factors will cancel in the ratios of path integrals that we are interested in, such as (\ref{gra}), (\ref{offw}) and (\ref{miqq}) below. The argument of the exponential in eqn. (\ref{bbkkp}) contains the action $S$ evaluated along the path $\mathbb{L}_{\alpha_1\alpha_2}$. Thus $\tilde a_{\alpha_1\alpha_2}$ is proportional to the probability amplitude for the particle to start at $q_{\alpha_1}$ and finish at $q_{\alpha_2}$. Let us consider
\begin{equation}
a_{\alpha_1\alpha_2}:=\frac{\tilde a_{\alpha_1\alpha_2}}{\vert\tilde a_{\alpha_1\alpha_2}\vert},
\label{gra}
\end{equation}
{\it i.e.}, the U(1)--valued phase of the path integral (\ref{bbkkp}). Now assume that $U_{\alpha_1}\cap U_{\alpha_2}$ is nonempty,
\begin{equation} 
U_{\alpha_1\alpha_2}:=U_{\alpha_1}\cap U_{\alpha_2}\neq \phi.
\label{neptyy}
\end{equation}
and define, for $q_{\alpha_{12}}\in U_{\alpha_1\alpha_2}$,
$$
\tau_{\alpha_1\alpha_2}\colon U_{\alpha_1\alpha_2}\longrightarrow {\rm U(1)}
$$
\begin{equation}
\tau_{\alpha_1\alpha_2}(q_{\alpha_{12}}):=a_{\alpha_1\alpha_{12}}(q_{\alpha_{12}})a_{\alpha_{12}\alpha_2}(q_{\alpha_{12}}).
\label{csgkdh}
\end{equation}
Thus $\tau_{\alpha_1\alpha_2}$ equals the U(1)--valued phase of the probability amplitude for the following transition: starting at $q_{\alpha_1}$, the particle reaches $q_{\alpha_2}$ {\it after}\/ traversing the variable midpoint $q_{\alpha_{12}}$. One can readily prove that (\ref{csgkdh}) qualifies as a gerbe trivialisation on $\mathbb{F}$. This trivialisation can be expressed as 
\begin{equation}
\tau_{\alpha_1\alpha_2}(q_{\alpha_{12}})=\frac{\tilde\tau_{\alpha_1\alpha_2}(q_{\alpha_{12}})}{\vert\tilde\tau_{\alpha_1\alpha_2}(q_{\alpha_{12}})\vert},
\label{offw}
\end{equation}
where
\begin{equation}
\tilde\tau_{\alpha_1\alpha_2}(q_{\alpha_{12}})\sim\int{\rm D}\mathbb{L}_{\alpha_1\alpha_2}(\alpha_{12})\exp\left[\frac{{\rm i}}{\hbar}\,S\left(\mathbb{L}_{\alpha_1\alpha_2}(\alpha_{12})\right)\right].
\label{beth}
\end{equation}
The functional integral (\ref{beth}) extends over all paths $\mathbb{L}_{\alpha_1\alpha_2}(\alpha_{12})$ that meet the requirements stated after eqn. (\ref{csgkdh}).

\subsection{The 2--cocycle}\label{dos}

Next consider three points and their respective charts
\begin{equation}
q_{\alpha_1}\in U_{\alpha_1},\qquad q_{\alpha_2}\in U_{\alpha_2}, \qquad q_{\alpha_3}\in U_{\alpha_3}
\label{ttppo}
\end{equation}
such that the triple overlap $U_{\alpha_1}\cap U_{\alpha_2}\cap U_{\alpha_3}$ is nonempty,
\begin{equation}
U_{\alpha_1\alpha_2\alpha_3}:=U_{\alpha_1}\cap U_{\alpha_2}\cap U_{\alpha_3}\neq \phi.
\label{pacca}
\end{equation}
Once the trivialisation (\ref{csgkdh}) is known, the 2--cocycle $g_{\alpha_1\alpha_2\alpha_3}$ defining a gerbe on $\mathbb{F}$ is given by 
$$
g_{\alpha_1\alpha_2\alpha_3}\colon U_{\alpha_1\alpha_2\alpha_3}\longrightarrow {\rm U(1)}
$$
\begin{equation}
g_{\alpha_1\alpha_2\alpha_3}(q_{\alpha_{123}}):=\tau_{\alpha_1\alpha_2}(q_{\alpha_{123}})\tau_{\alpha_2\alpha_3}(q_{\alpha_{123}})\tau_{\alpha_3\alpha_1}(q_{\alpha_{123}}),
\label{eptcg}
\end{equation}
where all three $\tau$'s on the right--hand side are, by definition, evaluated at the same variable midpoint  
\begin{equation}
q_{\alpha_{123}}\in U_{\alpha_1\alpha_2\alpha_3}.
\label{vef}
\end{equation}
Thus $g_{\alpha_1\alpha_2\alpha_3}$ equals the U(1)--phase of the probability amplitude for the following transition: starting at $q_{\alpha_1}$, the particle crosses $q_{\alpha_{123}}$ on its way to $q_{\alpha_2}$; from $q_{\alpha_2}$ it crosses $q_{\alpha_{123}}$ again on its way to $q_{\alpha_3}$; from $q_{\alpha_3}$ it traverses $q_{\alpha_{123}}$ once more before finally reaching $q_{\alpha_1}$. The corresponding closed path (see figure\footnote{Figure available upon request.}) is
\begin{equation}
\mathbb{L}_{\alpha_1\alpha_2\alpha_3}(\alpha_{123}):=\mathbb{L}_{\alpha_1\alpha_2}(\alpha_{123})+\mathbb{L}_{\alpha_2\alpha_3}(\alpha_{123})+\mathbb{L}_{\alpha_3\alpha_1}(\alpha_{123}).
\label{fuga}
\end{equation}
The 2--cocycle (\ref{eptcg}) can be expressed as the quotient
\begin{equation}
g_{\alpha_1\alpha_2\alpha_3}(q_{\alpha_{123}})=\frac{\tilde g_{\alpha_1\alpha_2\alpha_3}(q_{\alpha_{123}})}{\vert \tilde g_{\alpha_1\alpha_2\alpha_3}(q_{\alpha_{123}})\vert},
\label{miqq}
\end{equation}
where
\begin{equation}
\tilde g_{\alpha_1\alpha_2\alpha_3}(q_{\alpha_{123}})\sim\int{\rm D}\mathbb{L}_{\alpha_1\alpha_2\alpha_3}(\alpha_{123})\exp\left[\frac{{\rm i}}{\hbar}\,S\left(\mathbb{L}_{\alpha_1\alpha_2\alpha_3}(\alpha_{123})\right)\right].
\label{xam}
\end{equation}
The functional integral (\ref{xam}) extends over all paths $\mathbb{L}_{\alpha_1\alpha_2\alpha_3}(\alpha_{123})$ that meet the requirements stated after eqn. (\ref{vef}).

Given a closed loop $\mathbb{L}$, let $\mathbb{S}\subset\mathbb{F}$ be a 2--dimensional surface with boundary such that $\partial\mathbb{S}=\mathbb{L}$. By Stokes' theorem,
\begin{equation}
S(\mathbb{L})=\int_{\mathbb{L}}{\cal L}{\rm d}t=\int_{\partial\mathbb{S}}{\cal L}{\rm d}t=\int_{\mathbb{S}}{\rm d}{\cal L}\wedge{\rm d}t.
\label{toes}
\end{equation}
Any surface $\mathbb{S}$ such that $\partial\mathbb{S}=\mathbb{L}$ will satisfy eqn. (\ref{toes}) because the integrand ${\rm d}{\cal L}\wedge{\rm d}t$ is closed. Let us now choose $\mathbb{S}$ to bound a closed loop $\mathbb{L}_{\alpha_1\alpha_2\alpha_3}(\alpha_{123})$ as in eqn. (\ref{fuga}). Consider the first half of the leg $\mathbb{L}_{\alpha_1\alpha_2}(\alpha_{123})$, denoted $\frac{1}{2}\mathbb{L}_{\alpha_1\alpha_2}(\alpha_{123})$. The latter runs from $\alpha_1$ to $\alpha_{123}$. Consider also the second half of the leg $\mathbb{L}_{\alpha_3\alpha_1}(\alpha_{123})$, denoted $\frac{1}{2'}\mathbb{L}_{\alpha_3\alpha_1}(\alpha_{123})$, with a prime to remind us that it is the {\it second}\/ half: it runs back from $\alpha_{123}$ to $\alpha_1$ (see figure). The sum of these two half legs,
\begin{equation}
\frac{1}{2}\mathbb{L}_{\alpha_1\alpha_2}(\alpha_{123})+\frac{1}{2'}\mathbb{L}_{\alpha_3\alpha_1}(\alpha_{123}),
\label{mezzo}
\end{equation}
completes one roundtrip and it will, as a rule, enclose an area $\mathbb{S}_{\alpha_1}(\alpha_{123})$, unless the path from $\alpha_{123}$ to $\alpha_1$  happens to coincide exactly with the path from $\alpha_1$ to $\alpha_{123}$:
\begin{equation}
\partial\mathbb{S}_{\alpha_1}(\alpha_{123})=\frac{1}{2}\mathbb{L}_{\alpha_1\alpha_2}(\alpha_{123})+\frac{1}{2'}\mathbb{L}_{\alpha_3\alpha_1}(\alpha_{123}).
\label{soprano}
\end{equation}
Analogous conclusions apply to the other half legs $\frac{1}{2'}\mathbb{L}_{\alpha_1\alpha_2}(\alpha_{123})$, $\frac{1}{2}\mathbb{L}_{\alpha_3\alpha_1}(\alpha_{123})$, $\frac{1}{2}\mathbb{L}_{\alpha_2\alpha_3}(\alpha_{123})$ and $\frac{1}{2'}\mathbb{L}_{\alpha_2\alpha_3}(\alpha_{123})$ under cyclic permutations of 1,2,3 in the \v Cech indices $\alpha_1$, $\alpha_2$ and $\alpha_3$:
\begin{equation}
\partial\mathbb{S}_{\alpha_2}(\alpha_{123})=\frac{1}{2}\mathbb{L}_{\alpha_2\alpha_3}(\alpha_{123})+\frac{1}{2'}\mathbb{L}_{\alpha_1\alpha_2}(\alpha_{123}),
\label{tenor}
\end{equation}
\begin{equation}
\partial\mathbb{S}_{\alpha_3}(\alpha_{123})=\frac{1}{2}\mathbb{L}_{\alpha_3\alpha_1}(\alpha_{123})+\frac{1}{2'}\mathbb{L}_{\alpha_2\alpha_3}(\alpha_{123}).
\label{bajo}
\end{equation}
The boundaries of the three surfaces $\mathbb{S}_{\alpha_1}(\alpha_{123})$, $\mathbb{S}_{\alpha_2}(\alpha_{123})$ and $\mathbb{S}_{\alpha_3}(\alpha_{123})$ all pass through the variable midpoint $\alpha_{123}$, although we will no longer indicate this explicitly. We define their connected sum
\begin{equation}
\mathbb{S}_{\alpha_1\alpha_2\alpha_3}:=\mathbb{S}_{\alpha_1}+\mathbb{S}_{\alpha_2}+\mathbb{S}_{\alpha_3}.
\label{wdvdd}
\end{equation}
In this way we have
\begin{equation}
\mathbb{L}_{\alpha_1\alpha_2\alpha_3}=\partial\mathbb{S}_{\alpha_1\alpha_2\alpha_3}=
\partial\mathbb{S}_{\alpha_1}+\partial\mathbb{S}_{\alpha_2}+\partial\mathbb{S}_{\alpha_3}.
\label{bbvvdd}
\end{equation}
It must be borne in mind that $\mathbb{L}_{\alpha_1\alpha_2\alpha_3}$ is a function of the variable midpoint $\alpha_{123}\in U_{\alpha_1\alpha_2\alpha_3}$, even if we no longer indicate this explicitly.  Eventually one, two or perhaps all three of $\mathbb{S}_{\alpha_1}$, $\mathbb{S}_{\alpha_2}$ and $\mathbb{S}_{\alpha_3}$ may degenerate to a curve connecting the midpoint $\alpha_{123}$ with $\alpha_1$, $\alpha_2$ or $\alpha_3$, respectively. Whenever such is the case for all three surfaces, the closed trajectory $\mathbb{L}_{\alpha_1\alpha_2\alpha_3}$ cannot be expressed as the boundary of a 2--dimensional surface $\mathbb{S}_{\alpha_1\alpha_2\alpha_3}$. In what follows we will however exclude this latter possibility, so that at least one of the three surfaces on the right--hand side of (\ref{wdvdd}) does not degenerate to a curve.

In general we will not be able to compute the functional integral (\ref{xam}) exactly. However we can gain some insight from a steepest--descent approximation \cite{CH}, the details of which have been worked out in ref. \cite{PHASEGERBES}. We find
\begin{equation}
g^{(0)}_{\alpha_1\alpha_2\alpha_3}(q_{\alpha_{123}})=\exp\left[\frac{{\rm i}}{\hbar}S\left(\mathbb{L}^{(0)}_{\alpha_1\alpha_2\alpha_3}(\alpha_{123})\right)\right],
\label{vamp}
\end{equation}
the superindex ${}^{(0)}$ standing for {\it evaluation at the extremal}. The latter is that path which, meeting the requirements stated after eqn. (\ref{vef}),  minimises the action $S$. To summarise, by eqns. (\ref{xam}), (\ref{toes}), (\ref{wdvdd}), (\ref{bbvvdd}) and (\ref{vamp}), we can write the steepest--descent approximation to the 2--cocycle as
\begin{equation}
g_{\alpha_1\alpha_2\alpha_3}^{(0)}=\exp\left(\frac{{\rm i}}{\hbar}\int_{\mathbb{S}^{(0)}_{\alpha_1\alpha_2\alpha_3}}{\rm d}{\cal L}\wedge{\rm d}t\right),
\label{lecy}
\end{equation}
where $\mathbb{S}^{(0)}_{\alpha_1\alpha_2\alpha_3}$ is a minimal surface for the integrand ${\rm d}{\cal L}\wedge{\rm d}t$. We will henceforth drop the superindex ${}^{(0)}$, with the understanding that all our computations have been performed in the steepest--descent approximation.

\subsection{The connection}\label{nncc}

We can use eqns. (\ref{vamp}) and (\ref{lecy}) in order to compute the connection, at least to the same order of accuracy as the 2--cocycle itself. We find
\begin{equation}
A_{\alpha_1\alpha_2}=\frac{{\rm i}}{\hbar}\,\left({\cal L}\,{\rm d}t\right)_{\alpha_1\alpha_2},
\label{grag}
\end{equation}
\begin{equation}
B_{\alpha_2}-B_{\alpha_1}={\rm d}A_{\alpha_1\alpha_2}=\frac{{\rm i}}{\hbar}\,\left({\rm d}{\cal L}\wedge{\rm d}t\right)_{\alpha_1\alpha_2},
\label{mach}
\end{equation}
\begin{equation}
H\vert_{U_{\alpha}}={\rm d}B_{\alpha}.
\label{goed}
\end{equation}
A comment is in order. The potential $A$ is supposed to be a 1--form on configuration space $\mathbb{F}$, on which the gerbe is defined. As it stands in (\ref{grag}), due to the factor d$t$, $A$ is a 1--form on $\mathbb{F}\times\mathbb{R}$. If $\iota\colon\mathbb{F}\rightarrow\mathbb{F}\times\mathbb{R}$ denotes the natural inclusion, the 1--form $A$ in (\ref{grag}) is to be understood as its pullback $\iota^*({\cal L}{\rm d}t)$ onto $\mathbb{F}$. We will however continue to write it as ${\cal L}{\rm d}t$.

{}For a free particle, the right--hand side of eqn. (\ref{grag}) equals the pullback of the length element (\ref{kkaa}) to the geodesic wordline followed by the particle.
Then eqn. (\ref{mach}) implies that d$A$ is an area element. We can thus think of the 2--form $B$ as a locally--defined area element. The latter is subject to the gauge transformation law (\ref{mach}) across overlapping coordinate patches. In the presence of interactions, the Lagrangian ${\cal L}$ also includes a term of nongeometric origin (the potential energy $V(q)$) which apparently invalidates our interpretation of $B$ as a locally--defined area element. However, when integrating ${\cal L}$ along closed loops $\mathbb{L}=\partial\mathbb{S}$ as done here, the contribution from the potential energy cancels, and one may still regard the Neveu--Schwarz field $B$ as a surface element.

\section{Discussion}\label{kkbbn}

The machinery of quantum mechanics, as encoded in Feynman's exponential (\ref{bbmk}) of the classical action, has been neatly packaged in the geometric language of  gerbes on configuration space $\mathbb{F}$. In particular, this has produced a Neveu--Schwarz field $B$. 

We can now return to section \ref{intt} and prove the statement made there, namely, that the exchange (\ref{rchp}) is equivalent to a Heisenberg--like noncommutativity (\ref{cghp}) for the space coordinates. We first show that eqn. (\ref{rchp}) implies eqn. (\ref{cghp}). The starting point is that we can apply the exchange (\ref{rchp}) to our theory. We have in ref. \cite{ME} shown that this exchange is equivalent to the existence of a minimal length scale $L_P$ and, hence, to the existence of an area scale $a=L_P^2$. The latter arises as the result of integrating the 2--form $B$ over the smallest possible surface. Express the Neveu--Schwarz field $B$ in terms of de Rham--cohomology indices as $B_{\mu\nu}{\rm d}q^{\mu}\wedge{\rm d}q^{\nu}$. It has been known for long \cite{LANDAU} that a nonvanishing $B_{\mu\nu}$ across the $q^{\mu},q^{\nu}$ plane induces the space noncommutativity of eqn. (\ref{cghp}) with $\theta^{\mu\nu}=aB_{\mu\nu}^{-1}$, hence our statement follows. (We have assumed as usual that det$(\theta^{\mu\nu})\neq 0\neq$det$(B_{\mu\nu})$).

The converse also holds true, {\it i.e.}, eqn. (\ref{cghp}) implies eqn.  (\ref{rchp}). Here our starting point is the validity of eqn. (\ref{cghp}). Then the right--hand side of eqn. (\ref{cghp}) defines a Neveu--Schwarz field $B_{\mu\nu}=a(\theta^{\mu\nu})^{-1}$ on $\mathbb{F}$ which we can pick as a connection on a gerbe. Now the surface integral $\int_{\mathbb{S}} B$ is an area, for which there is an elementary quantum $a$. Then, by eqn. (\ref{rzf}), we have a quantum of length $L_P$. Next apply the arguments around eqns. (\ref{rkcak}) and (\ref{paddy}) in order to map the area $\int_{\mathbb{S}} B$ into $L_P^4/\int_{\mathbb{S}} B$.  Let us now, in eqns. (\ref{toes}) and (\ref{mach}), pick $\mathbb{S}$ such that $\mathbb{S}\subset U_{\alpha_1}\cap U_{\alpha_2}$, and set $B_{\alpha_1}=0$ by a local choice of gauge on $U_{\alpha_1}$.  Then we have $S(\mathbb{L})=-{\rm i}\hbar\int_{\mathbb{S}}B$. So we can write the exponential of $L_P^4/\int_{\mathbb{S}} B$ as in eqn. (\ref{bbmc}), and our assertion is proved.

We would finally like to stress the fact there is nothing wrong with a theory of {\it point}\/ particles (as opposed to {\it extended}\/ objects: strings, branes) giving rise to a Neveu--Schwarz 2--form $B_{\mu\nu}$, rather than the usual 1--form $A_{\mu}$. Surfaces, and therefore 2--forms, arise naturally as a consequence of considering {\it closed}\/ 1--dimensional paths. There is, however, a deeper reason for the appearance of the 2--form $B_{\mu\nu}$: the noncommutativity (\ref{cghp}) of the space coordinates. Under the latter, 1--dimensional paths become blurred into 2--dimensional surfaces. In turn, a nonvanishing right--hand side in eqn. (\ref{cghp}) follows from the requirement that gravity play its role, enforcing the existence of a minimal length scale $L_P$. 

{\bf Acknowledgements}

It is a great pleasure to thank Albert--Einstein--Institut (Potsdam, Germany) for hospitality during the preparation of this article. This work has been supported by Ministerio de Educaci\'{o}n y Ciencia (Spain) through grant FIS2005--02761, by Generalitat Valenciana, by EU FEDER funds, by EU network MRTN--CT--2004--005104 ({\it Constituents, Fundamental Forces and Symmetries of the Universe}), and by Deutsche Forschungsgemeinschaft.


\begin{thebibliography}{99}


\bibitem{LANDAU}
L. Landau and E. Lifshitz, {\it Quantum Mechanics}, vol. 3 of {\it Course of Theoretical Physics}, Butterworth--Heinemann, Oxford (2000).

\bibitem{NCG}
G. Landi, {\it An Introduction to Noncommutative Spaces and their Geometry},
Lecture Notes in Physics {\bf 51}, Springer, Berlin (1997);\\
J. Madore, {\it An Introduction to Noncommutative Geometry and its Physical 
Applications}, London Mathematical Society Lecture Note Series {\bf 257}, 
Cambridge University Press, Cambridge (1999);\\
for a review see, {\it e.g.},\\
R. Szabo, {\it Int. J. Mod. Phys.} {\bf A19} (2004) 1837.

\bibitem{PADUNO}
T. Padmanabhan, {\it Phys. Rev. Lett.}  {\bf 78} (1997) 1854;  {\it Phys. Rev.} {\bf D57}  (1998) 6206.

\bibitem{ME}
J.M. Isidro, {\it Mod. Phys. Lett.} {\bf A20} (2005) 2913; {\tt hep-th/0507150}.

\bibitem{AZCA}
J. de Azc\'arraga and J. Izquierdo, {\it Lie Groups, Lie Algebras, 
Cohomology and some Applications in Physics}, Cambridge University Press, 
Cambridge (1995).

\bibitem{GERBES}
J. Brylinski, {\it Loop Spaces, Characteristic Classes and Geometric Quantization}, Progress in Mathematics {\bf 107}, Birkh\"auser, Boston (1993).

\bibitem{PINZULSTERN}
A. Pinzul and A. Stern, {\tt hep-th/0402220}.

\bibitem{RECENT}
L. Breen and W. Messing, {\tt math.AG/0106083};\\
M. Stern, {\tt hep-th/0209192};\\
M. Caicedo, I. Mart\' in and A. Restuccia, {\it Annals Phys.} {\bf 300} (2002) 32;\\
H. Pfeiffer, {\it Annals Phys.} {\bf 308} (2003) 447;\\
A. Carey {\it et al.}, {\tt math.DG/0410013};\\
V. Mathai and D. Roberts, {\tt hep-th/0509037};\\
P. Aschieri, L. Cantini and B. Jur\v co, {\it Comm. Math. Phys.} {\bf 254} (2005) 367;\\
J. Kalkkinen, {\it Fortsch. Phys.} {\bf 53} (2005) 913;\\
J. Mickelsson, {\tt math-ph/0603031}.

\bibitem{VARIOUS}
G. Bertoldi, A. Faraggi and M. Matone,  {\it Class. Quant. Grav.} {\bf 17} (2000) 3965;\\
H. Garc\'\i a--Compe\'an, O. Obreg\'on, C. Ram\'\i rez and M. Sabido, {\it Phys. Rev.} {\bf D68} (2003) 045010;\\
D. Minic and C. Tze, {\tt hep-th/0309239};\\
J.M. Isidro, {\tt hep-th/0411015}.

\bibitem{VAFA}
C. Vafa, {\tt hep-th/9702201}.

\bibitem{PORRATI}
T. Giveon, M. Porrati and E. Rabinovici, {\it Phys. Rep.} {\bf 244} (1994) 77.

\bibitem{ORLANDO}
O. Alvarez, {\it Comm. Math. Phys.} {\bf 100} (1985) 279.

\bibitem{PICKEN}
R. Picken, {\it A Cohomological Description of Abelian Bundles and Gerbes}, in {\it Twenty Years of Bialowieza: a Mathematical Anthology}, 217--228, World Sci. Monogr. Ser. Math. {\bf 8}, World Scientific Publications, Hackensack NJ (2005).

\bibitem{CH}
R. Courant and D. Hilbert, {\it Methods of Mathematical Physics}, Wiley, New York (1989).

\bibitem{PHASEGERBES}
J.M. Isidro, {\tt hep-th/0512241}.


\end{thebibliography}
\end{document}